\documentclass[a4paper]{jpsj-suppl}
\usepackage{txfonts}

\newcommand{\dpt}{{\delta\it{p_\mathrm{T}}}}
\newcommand{\dat}{{\delta\it{\alpha_\mathrm{T}}}}
\newcommand{\dphit}{{\delta\it{\phi_\mathrm{T}}}}
\newcommand{\mupt}{{\it{p}^\ell_\mathrm{T}}}
\newcommand{\tdpt}{$\dpt$ }
\newcommand{\tdat}{$\dat$ }
\newcommand{\tdphit}{$\dphit$ }
\newcommand{\tmupt}{$\mupt$}

\newcommand{\qsq}{Q^{2}}
\newcommand{\tqsq}{$\qsq$ }

\newcommand{\etrans}{\omega}
\newcommand{\tetrans}{$\etrans$ }

\newcommand{\hadrmass}{W}
\newcommand{\thadrmass}{$\hadrmass$ }

\newcommand{\figtxt}{Fig.~}

\title{Examining nuclear effects in neutrino interactions with transverse
kinematic imbalance}

\author{Luke \textsc{Pickering}$^{1}$}
\inst{$^{1}$Imperial College London}
\email{luke.pickering08@imperial.ac.uk}
\recdate{January 31, 2016}

\abst{We present a Monte Carlo truth study examining nuclear effects in
charged-current neutrino interactions using observables constructed in the
transverse plane.
Three distributions are introduced that show very weak dependence on neutrino
flux and its associated uncertainty.
Measurements comparing these distributions between
quasi-elastic-like and single charged pion final states will provide new
constraints of nuclear effects.
It is suggested that the on-axis position in the NuMI beam provides the correct
flux to take advantage of this reduced energy dependence in measuring nuclear
effect-generated transverse imbalances.}

\kword{neutrino-nucleus interaction, neutrino energy dependence, nuclear
effects, transverse plane}

\begin{document}
\maketitle

\section{Introduction}

Correctly reconstructing the energy of a neutrino interaction is an important
part of both neutrino oscillation and neutrino scattering physics.
Using nuclear targets, as opposed to free nucleon targets,
obfuscates the neutrino interaction and imparts significant uncertainties and
model dependencies onto the reconstructed neutrino energy.
Current neutrino beams are wide-band and the spread in event-by-event neutrino
energy is significant, thereby making the impact of nuclear effects difficult
to isolate.
Understanding the hadronic system is key for investigations of nuclear
effects; it holds all available information about the type of interaction that
occurred.

This talk presents a Monte Carlo truth study into the use of variables defined
in the plane transverse to the incoming neutrino.
In the absence of nuclear effects, the transverse momentum in the final state
should be identically balanced.
Therefore, nuclear effects that affect the kinematics of final state particles
can be seen as deviations from finely balanced systems.
For more details the reader is directed to \cite{tttpaper}.

This study mainly examines predictions from the NuWro generator \cite{NuWro}.
However, NEUT \cite{NEUT}, GENIE \cite{GENIE}, and GiBUU \cite{GiBUU} have also
been investigated and the procedures used are fully generator-independent.

\section{Energy Transfer and Hadronic System Energy in QE and RES Interactions}

\begin{figure}
\centering
\includegraphics[width=0.49\textwidth]{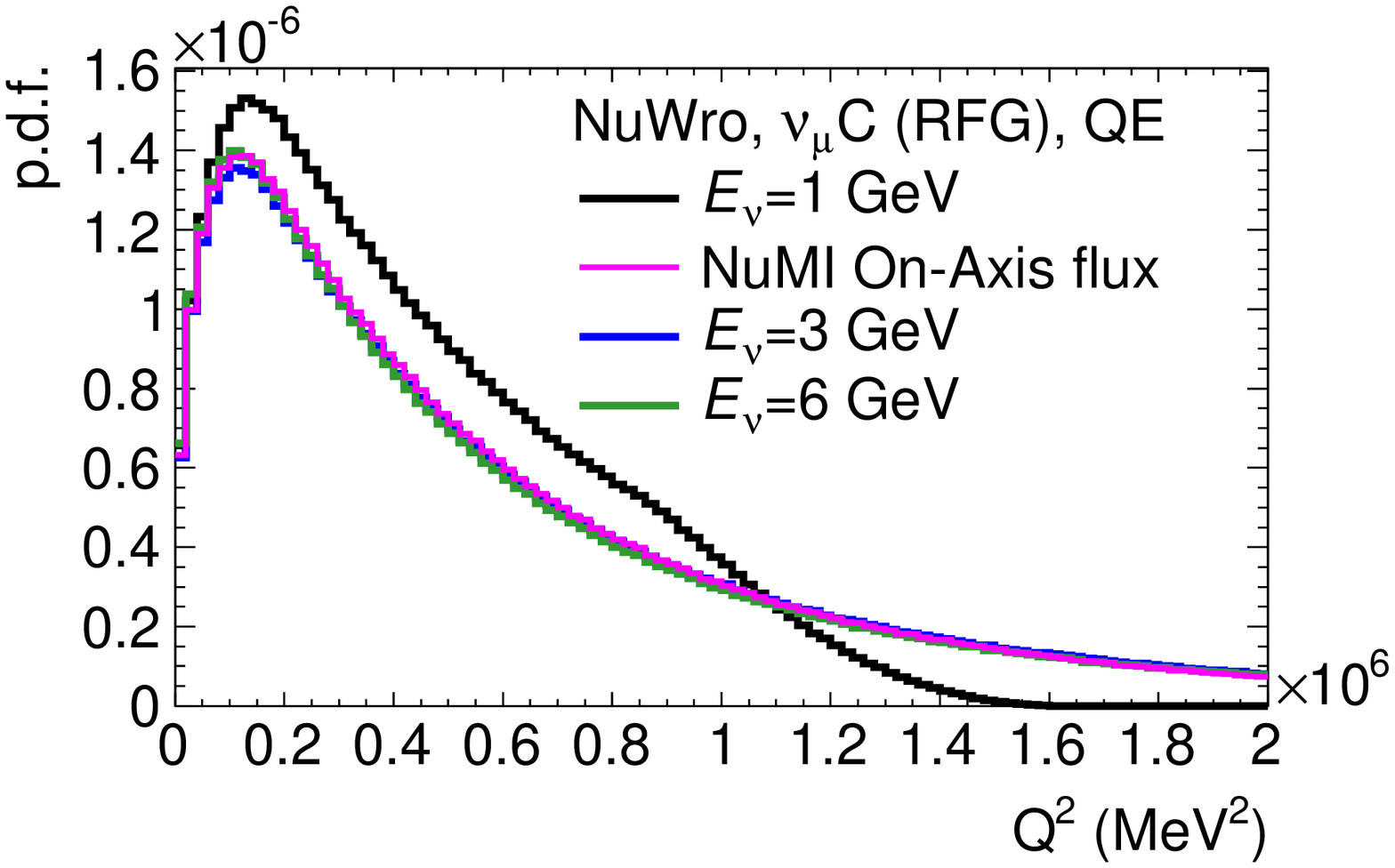}
\includegraphics[width=0.49\textwidth]{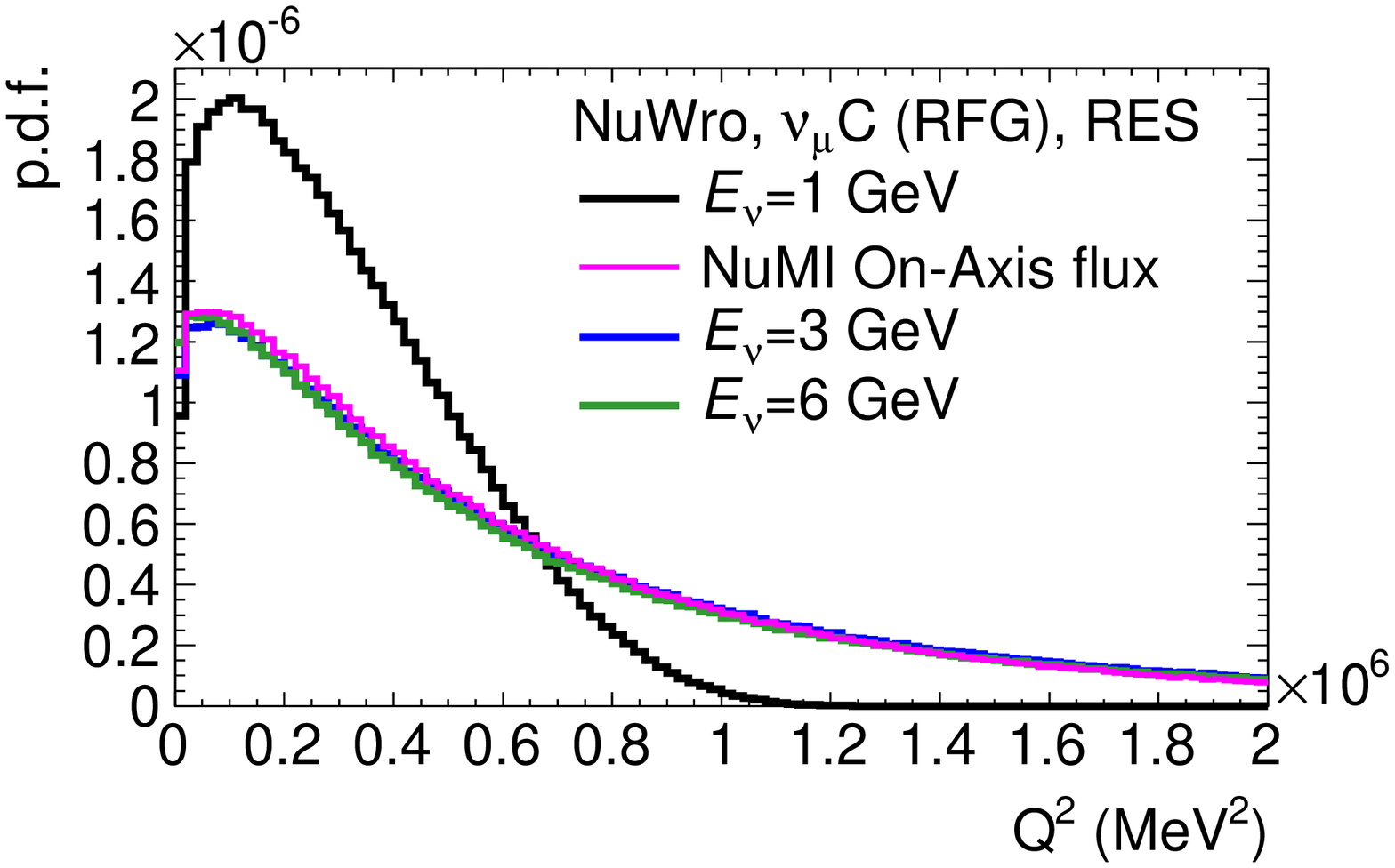}
\caption{Four momentum transfer for QE (Llewellyn Smith
\cite{LlewellynSmith}) and RES (Rein Sehgal \cite{ReinSehgal})
interactions as predicted by NuWro.
The \tqsq shape is asymptotically independent of the neutrino energy above some
threshold, leading to decoupling of the neutrino energy and the energy of the
final state hadronic system.
The NuMI on-axis beam is sufficiently high energy to cause this decoupling in
both the QE and RES channels.}
\label{fig:Q2SatQE}
\end{figure}

For quasi-elastic (QE \cite{LlewellynSmith}) and resonant
pion production (RES \cite{ReinSehgal})
neutrino interactions the phase space for four momentum transfer, \tqsq, is
bounded.
At larger four momentum transfer, the mediating particle scatters off quarks
and results in so called deep inelastic scatters.
\figtxt\ref{fig:Q2SatQE} shows the NuWro predicted event
rate for QE and RES interactions, as a function of \tqsq for a number of
different neutrino fluxes.
For neutrino energies $\lesssim2$ GeV, the \tqsq shape changes
significantly with increasing neutrino energy.
However, the shape change between $3$ GeV and $6$ GeV is minimal.
For QE and RES interactions at these energies, the models predict that almost
all of the possible \tqsq phase space for that interaction type is kinematically
accessible.
Predicted event rates for interactions in a real beam with a wide spectral
shape and peaked at sufficiently high energy, such as the NuMI on-axis beam
\cite{NUMIFLUX}, also exhibit the same \tqsq shape.

In a neutrino interaction where a single hadronic particle, with
invariant hadronic mass \thadrmass, is excited from an initial state nucleon,
$N$, the energy transfer, \tetrans, can be written as

\begin{align*}
  \etrans = \frac{\qsq + \hadrmass^2 - m^2_N}{2\sqrt{m^2_N+p^2_N}},
\end{align*}
where $m_N$ and $p_N$ are the mass and three momentum of the initial state
nucleon.
For QE interactions, $\nu_\mu + n \rightarrow \mu + p$, \thadrmass is the proton
mass.
For RES interactions, hadronic mass cuts that isolate single pion production via
a delta resonance are frequently used, \textit{e.g.} in MINERvA charged
current single pion production \cite{minervacc1pi}.
This results in a hadronic mass distributed about $1.2$ GeV.
The initial state nucleon momentum distribution is separate from the neutrino
interaction.
The only remaining unknown is \tqsq.
As discussed, above some threshold energy, \tqsq is largely independent of the
neutrino energy.
Therefore, in an appropriate neutrino beam, the event-by-event energy transfer
to the hadronic system should be largely decoupled from the neutrino energy.
The on-axis position in the NuMI beam has a peak energy high enough to cause
such decoupling for QE and for RES events.
The following discussions focus on NuWro predicted event rate shapes for a
carbon target exposed to the NuMI on-axis muon neutrino flux.

\section{Transverse Variables in QE Interactions}
\subsection{Definition}

\begin{figure}
\centering
\includegraphics[width=0.7\textwidth]{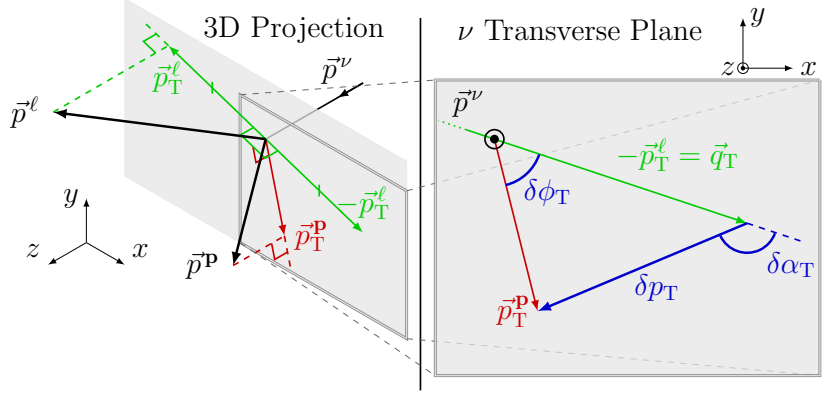}
\caption{A schematic definition of three transverse variables, \tdpt,
\tdat, and \tdphit, in a QE muon neutrino interaction.}
\label{fig:stvdef}
\end{figure}

Fig.~\ref{fig:stvdef} shows the schematic definitions of three variables defined
in the plane transverse to the incoming neutrino, \tdpt, \tdat, and \tdphit.

These three variables are defined as:

\begin{align*}
  \delta\vec{p}_\mathrm{T} &= \vec{p}^{\ell}_\mathrm{T} + \vec{p}^{p}_\mathrm{T},\\
  \dphit &= \arccos\frac{-\vec{p}^{\ell}_T \cdot \vec{p}^{N}_\mathrm{T} } { p^{\ell}_\mathrm{T}p^{N}_\mathrm{T}},\ \textrm{and}\\
  \dat &= \arccos\frac{-\vec{p}^{\ell}_T \cdot \delta\vec{p}_\mathrm{T}}{p^{\ell}_\mathrm{T}\dpt}.
\end{align*}

\tdpt is the overall three momentum imbalance in the transverse plane.
In the absence of nuclear effects---and detector effects---\tdpt should be
described by a Dirac delta function at $\dpt = 0$.
The transverse momentum imbalance has been investigated in neutrino scattering
previously.
The Big European Bubble Chamber examined transverse momentum
imbalance in deuterium exposed to wide band neutrino beams at the CERN SPS
\cite{bebbal}.
Their data set is significantly dominated by deep inelastic scattering events
due to the beam energy.
NOMAD used transverse kinematic imbalance as a selection variable
for QE events, also in a higher energy beam than used by current neutrino
experiments (which require $\mathcal{O}\left(\textrm{GeV}\right)$ beams for use
in neutrino oscillation physics) \cite{nomadqe}.

\tdphit describes the angular difference between the observed
final state and the case where the final state particles are back-to-back in
the transverse plane.
A highly balanced system is characterised by $\dphit=0$.
\tdphit has been used in QE event classifiers at NOMAD and INGRID
\cite{nomadqe,ingridqe}.
The MINERvA collaboration measured the \textit{coplanarity angle},
$\varphi = \pi - \dphit$, in their `CC with no pions in the final state' event
selection and find their data to largely agree with a GENIE simulation \cite{minervadphit}.

\tdat can be used to classify a process as having an `accelerating' or
`decelerating' effect on the hadronic system.
For $\dat<90^\circ$ the proton transverse momentum appears larger than expected
from the observed muon transverse momentum---an apparently accelerating
effect--and it appears less than expected for $\dat>90^\circ$.
\tdat was proposed and investigated for the first time in a recent truth study
\cite{tttpaper}.

Monte Carlo predictions for these three variables are presented in the following
sections along with discussions of features predicted by individual nuclear
effects.

\subsection{Energy dependence of the transverse variables}

\begin{figure}
\centering
\includegraphics[width=0.45\textwidth]{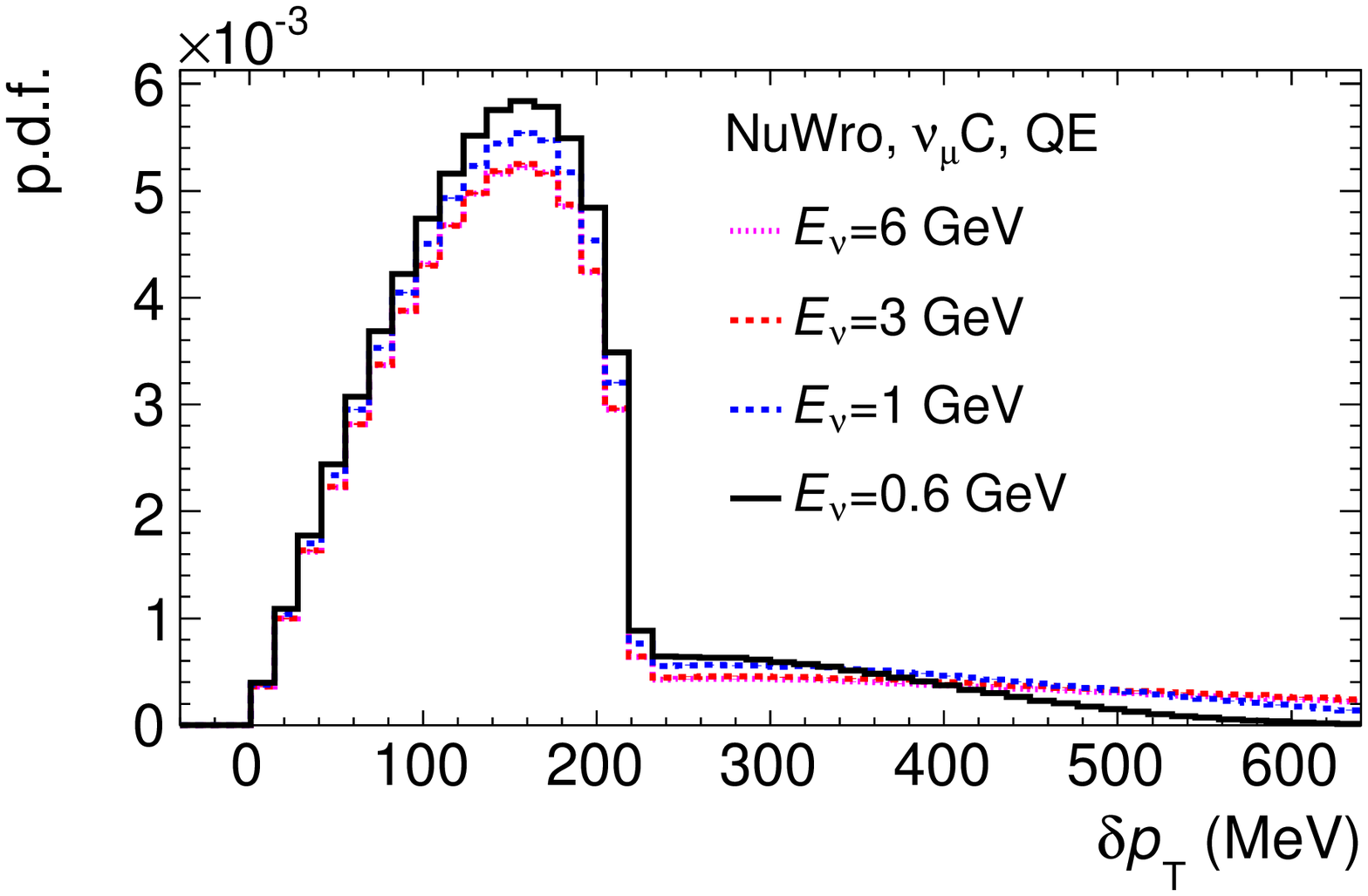}
\includegraphics[width=0.45\textwidth]{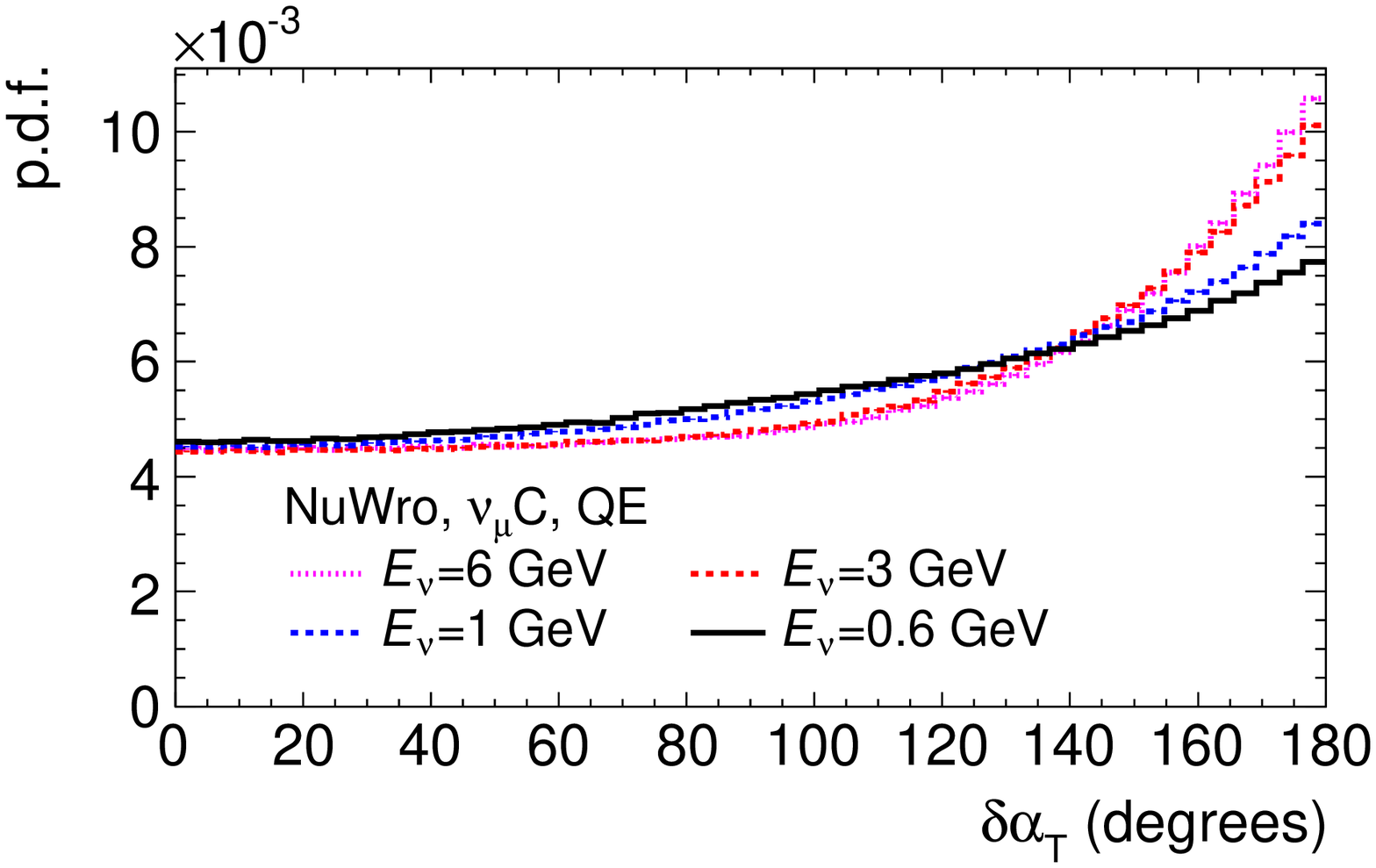}

\caption{Both \tdpt, (\textit{left}), and \tdat, (\textit{right}), exhibit
less neutrino energy dependence than final state lepton kinematics.}
\label{fig:dptdatscale}
\end{figure}

\tdpt is directly generated by nuclear effects. The direction of this transverse
momentum imbalance with respect to an axis defined by the charged lepton defines
\tdat. Fig.~\ref{fig:dptdatscale} shows how, over an order of magnitude of
beam energies, the change in the shape of \tdpt and \tdat is minimal.
This is not the case for charged lepton final state kinematics alone.
Such energy independence is desirable when attempting to isolate the effects of
using a nuclear target from the effects of a wide-band neutrino beam.

\begin{figure}
\centering

\begin{minipage}[]{0.4\textwidth}
\includegraphics[width=\textwidth]{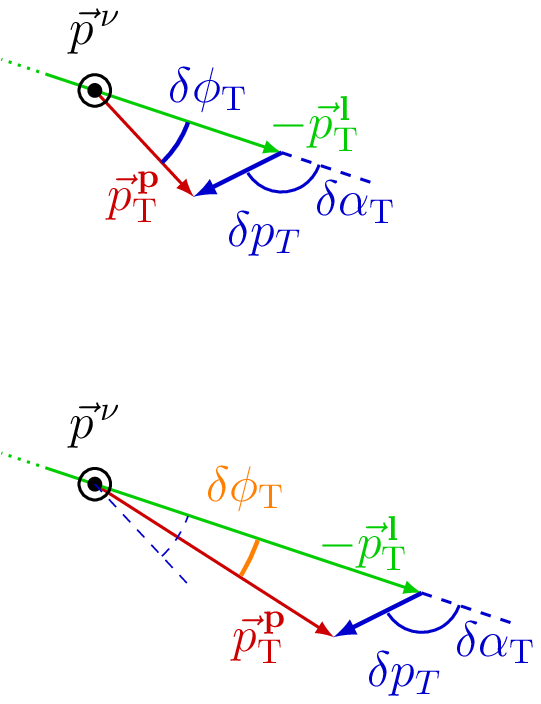}
\end{minipage}
\begin{minipage}[]{0.45\textwidth}
\includegraphics[width=\textwidth]{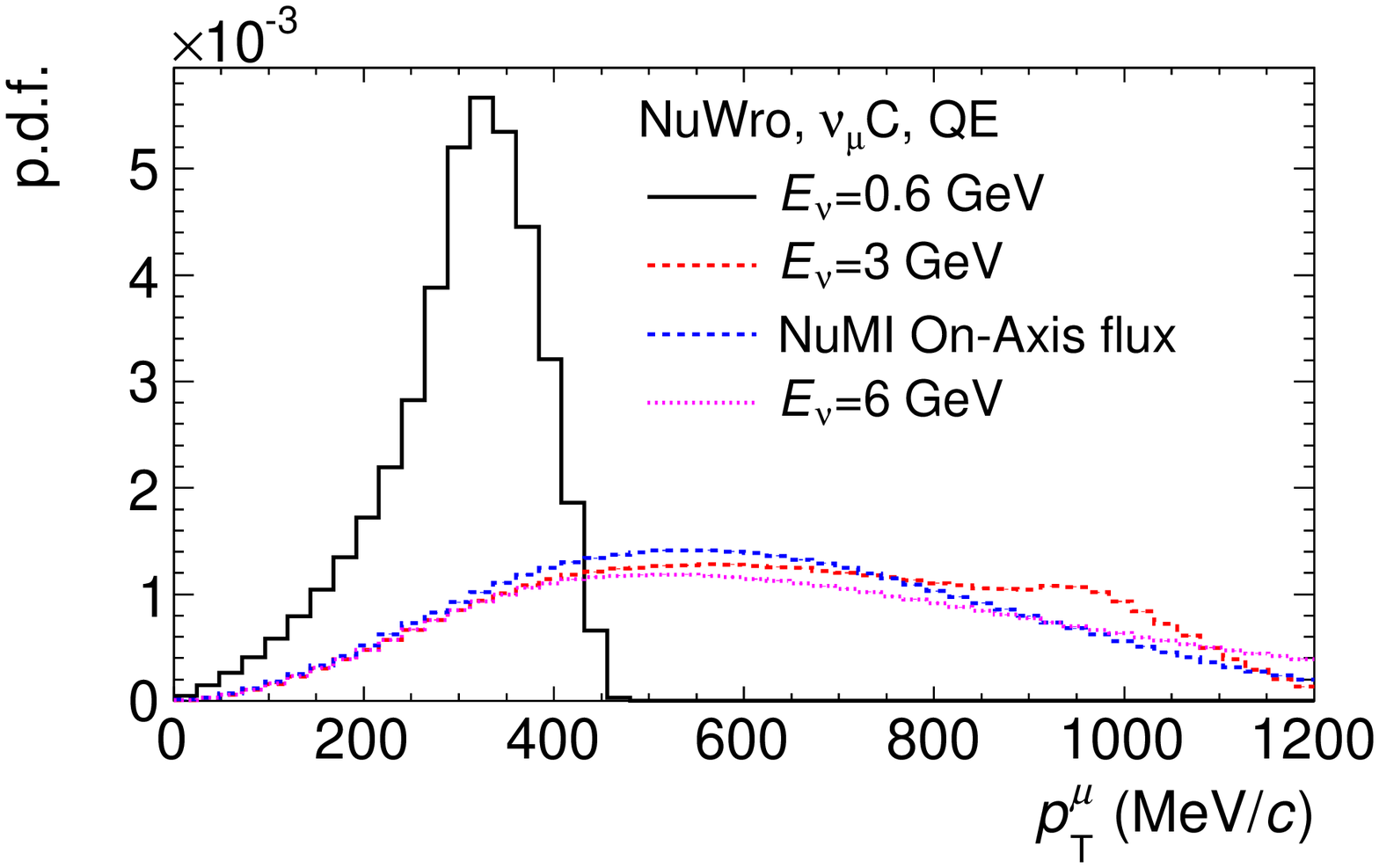}
\includegraphics[width=\textwidth]{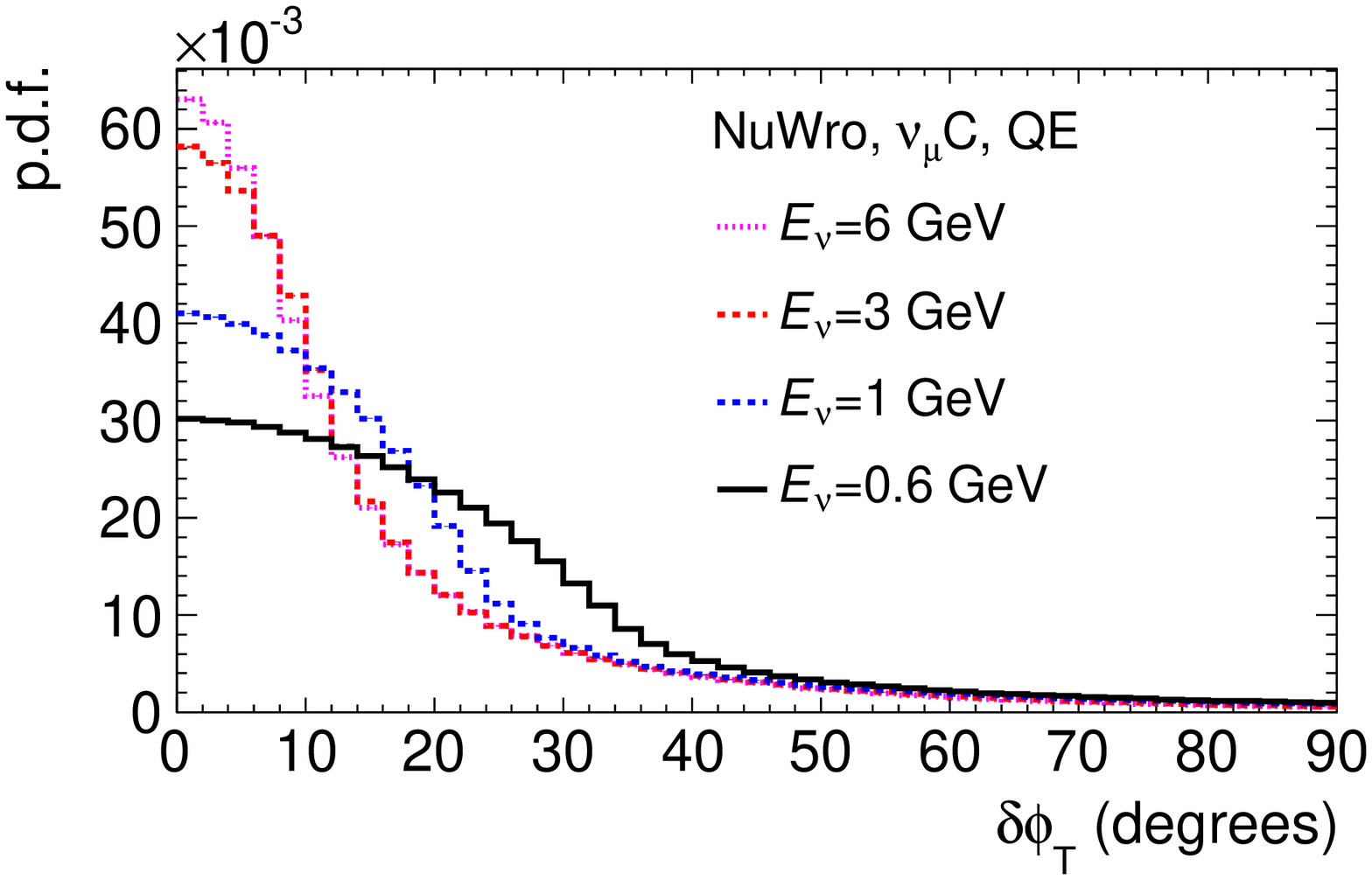}
\end{minipage}

\caption{(\textit{left}): For a given \tdpt and \tdat, which are generated by
distinct nuclear effects independent of the charged lepton kinematics,
the observed \tdphit depends on the magnitude of \tmupt.
(\textit{top right}): The \tmupt distribution varies strongly as a function of
neutrino energy. (\textit{below left}): For more energetic neutrinos, \tdphit
has a tighter peak due to the distribution of \tmupt, however, FSIs
result in a broader peak.}
\label{fig:dphitscale}
\end{figure}

However, the shape of \tdphit is more strongly influenced by the neutrino
energy.
This is because of the trigonometric dependence on the magnitude of the muon
transverse momentum.
Fig.~\ref{fig:dphitscale} shows how for a given \tdpt and \tdat, the resulting
\tdphit depends more strongly on the kinematics of the neutrino interaction
rather than just on the largely decoupled hadronic system.
When naively examining \tdphit, flux uncertainties are still more
conflated with FSI uncertainties.
One way to account for this dependence is to perform a double differential
measurement of $\frac{d^2\sigma}{d\dpt{}dp^{\ell}_T}$.

\subsection{Transverse Variables as Probes of Nuclear Effects}

\subsubsection{Fermi motion}

\begin{figure}
\centering
\includegraphics[width=0.49\textwidth]{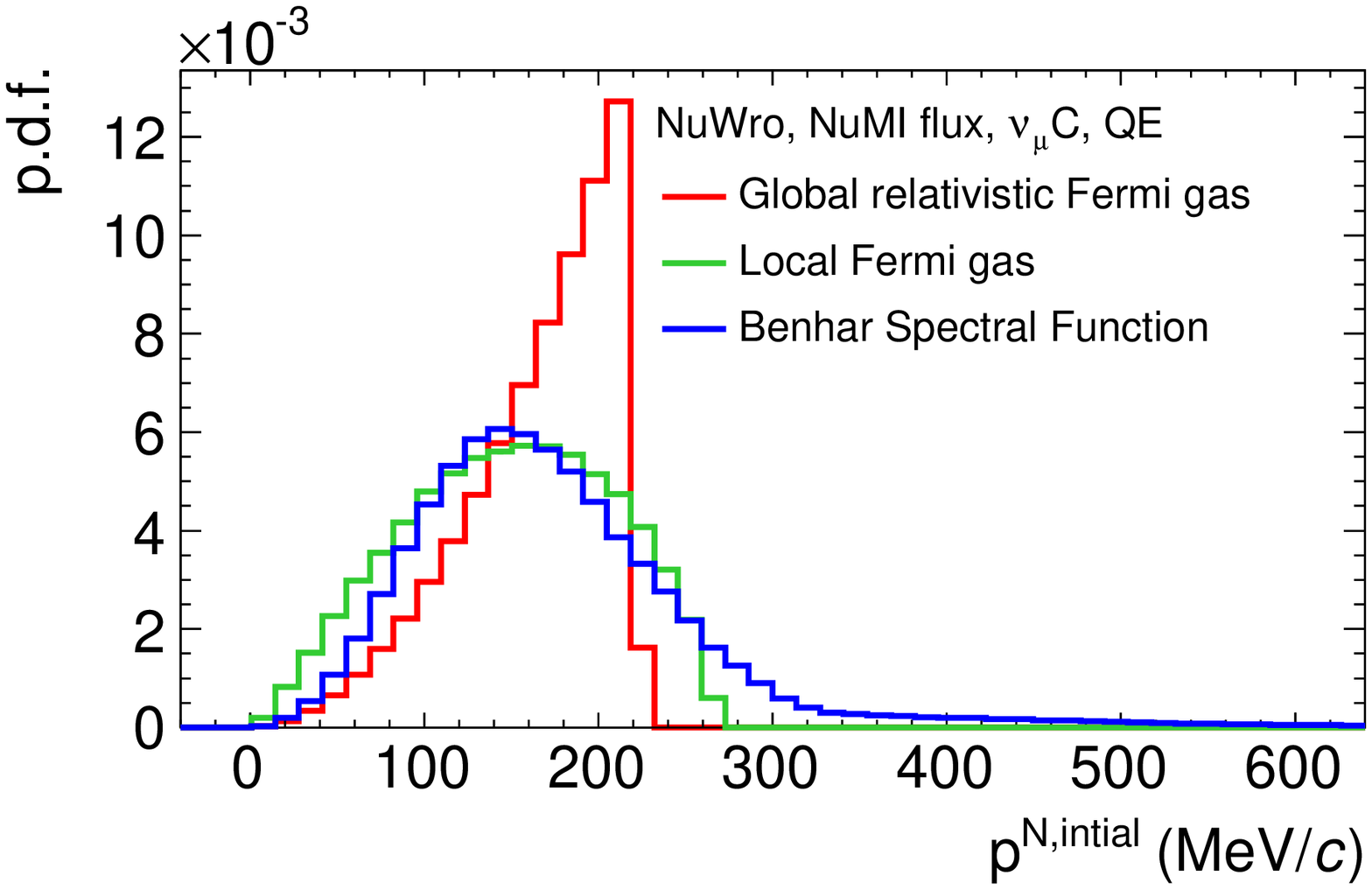}
\includegraphics[width=0.49\textwidth]{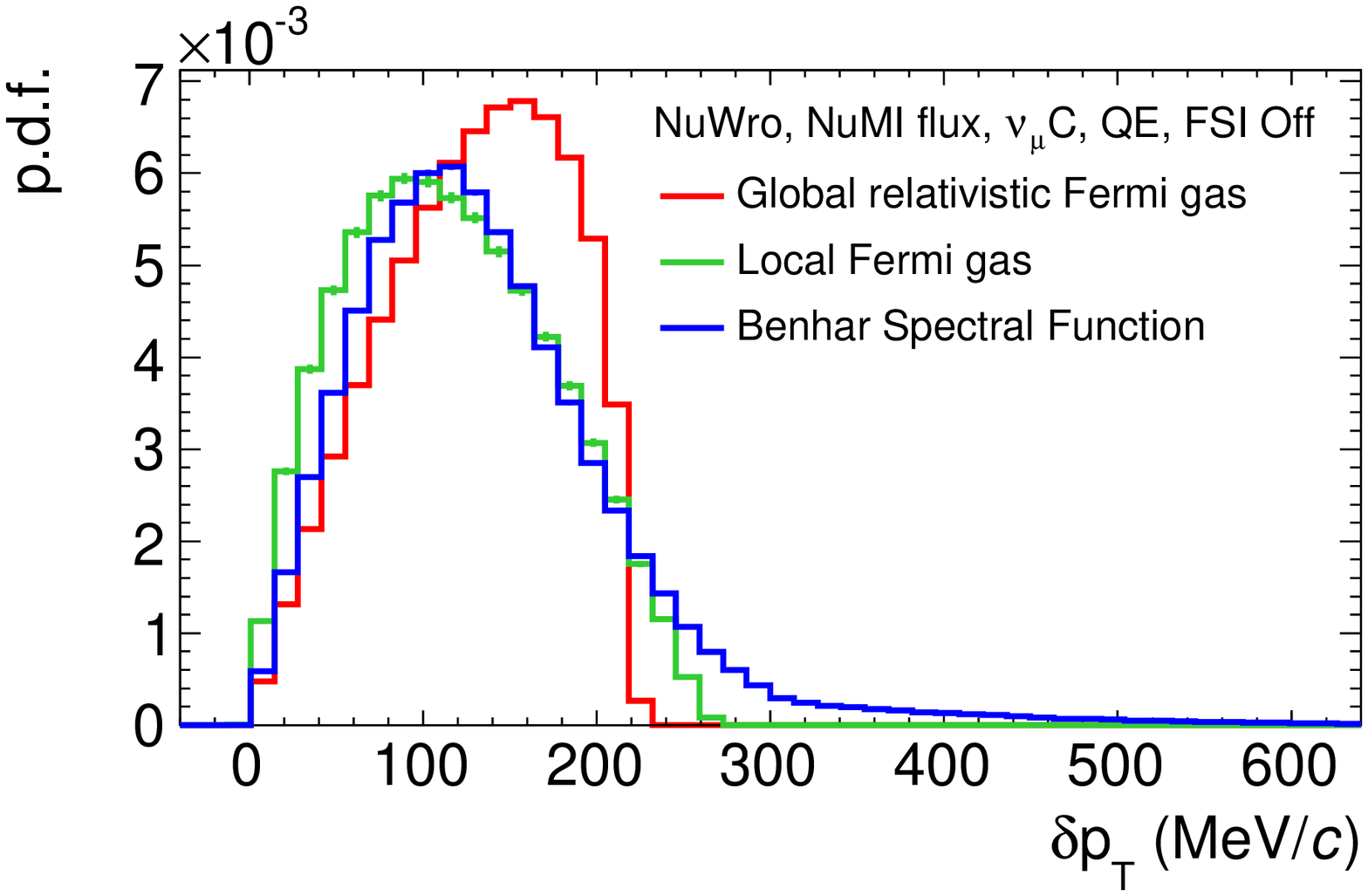}
\caption{(\textit{left}) The magnitude of the nucleon momentum distribution for
three widely used nuclear models (global relativistic Fermi gas (RFG), local Fermi gas (LFG), and Benhar spectral function (SF) \cite{BenharSF}).
(\textit{right}) The apparent transverse momentum resulting from the nucleon
Fermi motion, for QE interactions.}
\label{fig:initstate}
\end{figure}

When neutrinos scatter from free nucleons at rest, it is expected that the
transverse momentum of the final state particles identically balance:
$\dpt=0,\dphit=0,\dat \textrm{is undefined}$.
In a nuclear environment the nucleons are undergoing random Fermi motion which
imparts an unknowable event-by-event boost to the interacting system.
This boost introduces a characteristic shape to each of the three distributions.

In the absence of any Final State Interactions (FSIs) the apparent momentum
imbalance in the lab frame is generated purely by the Fermi motion of the
struck nucleon.
\tdpt is therefore distributed according to the transverse component of the
Fermi motion.
\figtxt\ref{fig:initstate} shows how differences in the nuclear model are
reflected in lab-frame \tdpt distributions.

\begin{figure}
\centering
\includegraphics[width=0.49\textwidth]{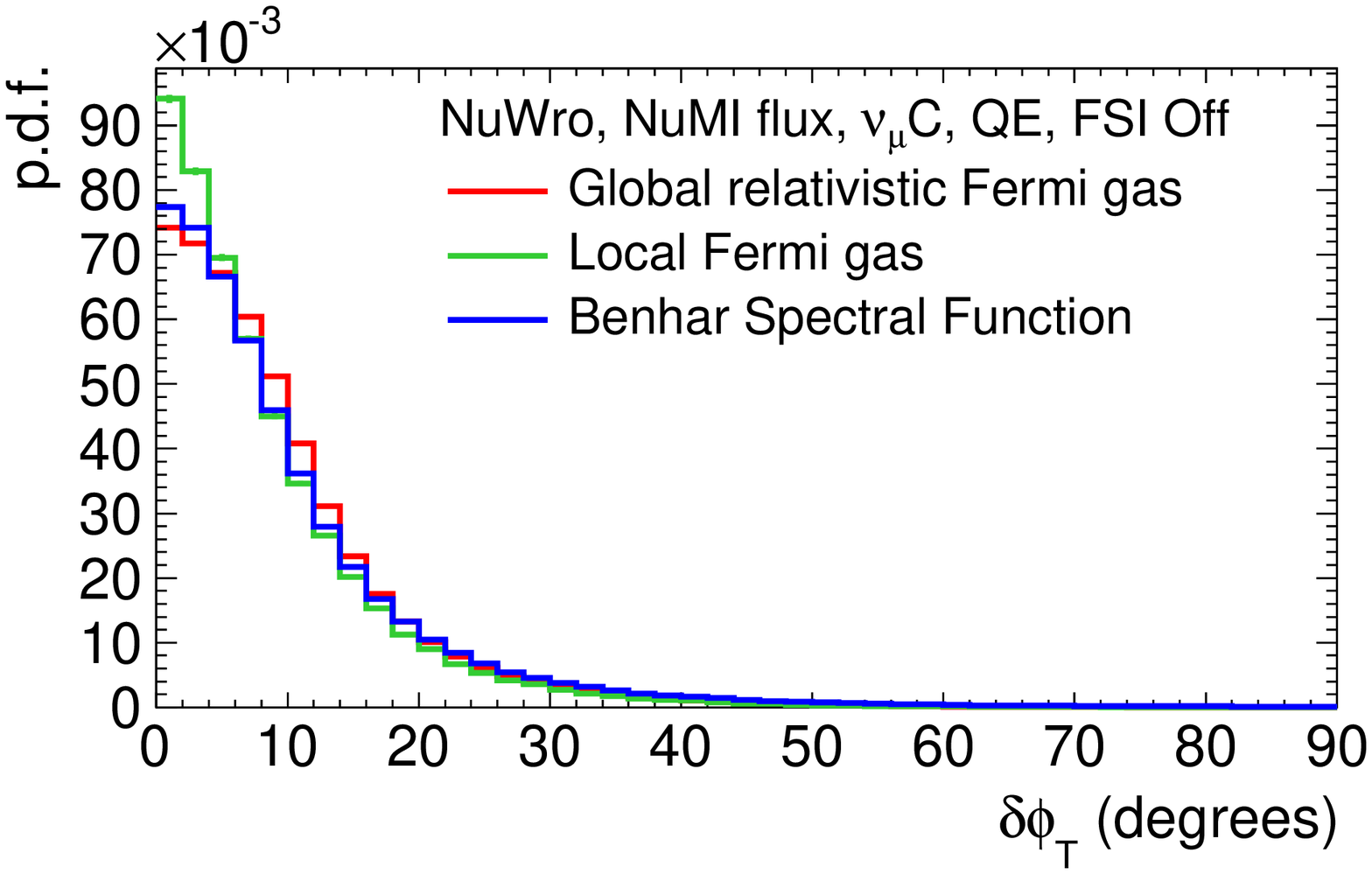}
\includegraphics[width=0.49\textwidth]{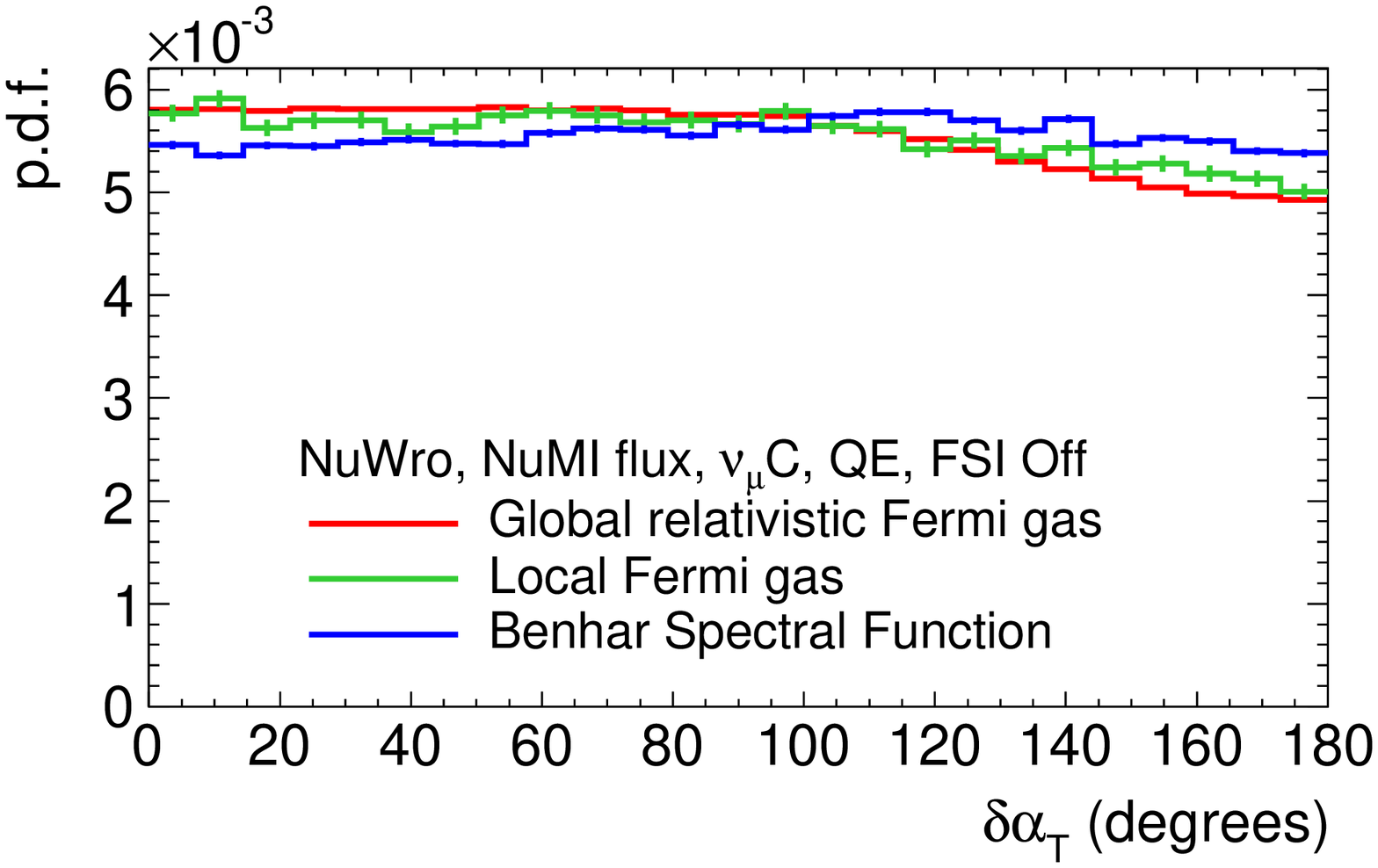}
\caption{The \tdphit (\textit{left}), and \tdat (\textit{right}),
predictions for three nuclear models in the absence of FSIs.}
\label{fig:initstate_datphit}
\end{figure}

\figtxt\ref{fig:initstate_datphit} shows the `FSI-off' predictions for \tdphit
and \tdat.
\tdphit gains some width as a result of the unknown boost; the
distribution is strongly peaked around $\dphit=0$.
\tdat can be defined for any transverse momentum imbalance $\dpt \neq 0$.
However, we expect the Fermi motion to be isotropic resulting a flat
distribution for the apparent acceleration or deceleration of the proton.

\subsubsection{Final State Interactions}

\begin{figure}
\centering
\includegraphics[width=0.6\textwidth]{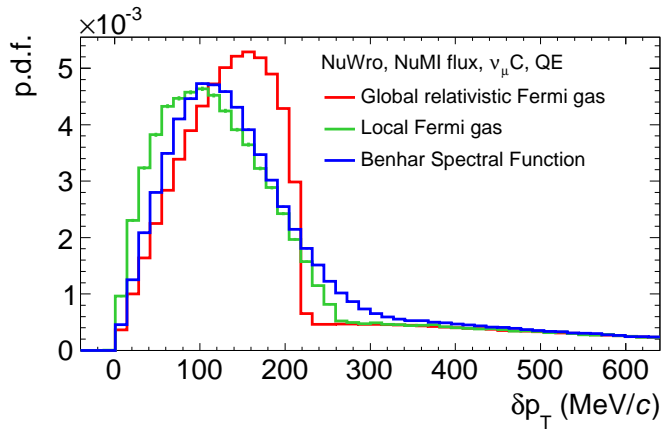}
\caption{The fully simulated transverse momentum imbalance for three
nuclear models.
Low \tdpt is dominated by the transverse projection of the nucleon momentum
distribution, as in \figtxt\ref{fig:initstate}.
At larger transverse momemtum imbalances, the \tdpt distribution becomes
largely independent of the initial state because of FSI smearing.}
\label{fig:initstateFSI}
\end{figure}

FSIs can refer to any re-scattering of the final hadronic system as it leaves the
nucleus.
This includes: elastic and inelastic re-scatters, absorption, charge exchange,
or spectator nucleon knock-out.
These processes all act to obscure the hadronic system `memory' of the
interaction.

\figtxt\ref{fig:initstateFSI} shows the \tdpt distributions for the three nuclear
models.
For $\dpt$ less than the Fermi momentum, $\textrm{k}_\textrm{f}$, the
distribution is almost entirely governed by the Fermi motion and for
$\dpt \gtrsim 2\textrm{k}_\textrm{f}$
are independent of the initial state and instead indicative of the FSI model.
The \tdphit distribution continues to broaden when FSI effects are enabled in
the simulation.

Herein the global relativistic Fermi gas (RFG) model will be used as the
nuclear model for the sake of simplicity.
The other nuclear models have been investigated and the results are
qualitatively similar.
The RFG model is very simple, and not the most realistic model.
When analyzing measured data, the most appropriate nuclear model should be used.

It might be naively expected that FSI processes generally transfer energy from
the vertex-exiting hadronic system to the nuclear environment.
The simulated distributions agree with these expectations, showing that FSIs
cause a pile-up of events at $\dat \sim 180^\circ$---characteristic of an
apparently decelerated hadronic system.

\begin{figure}
\centering
\includegraphics[width=0.51\textwidth]{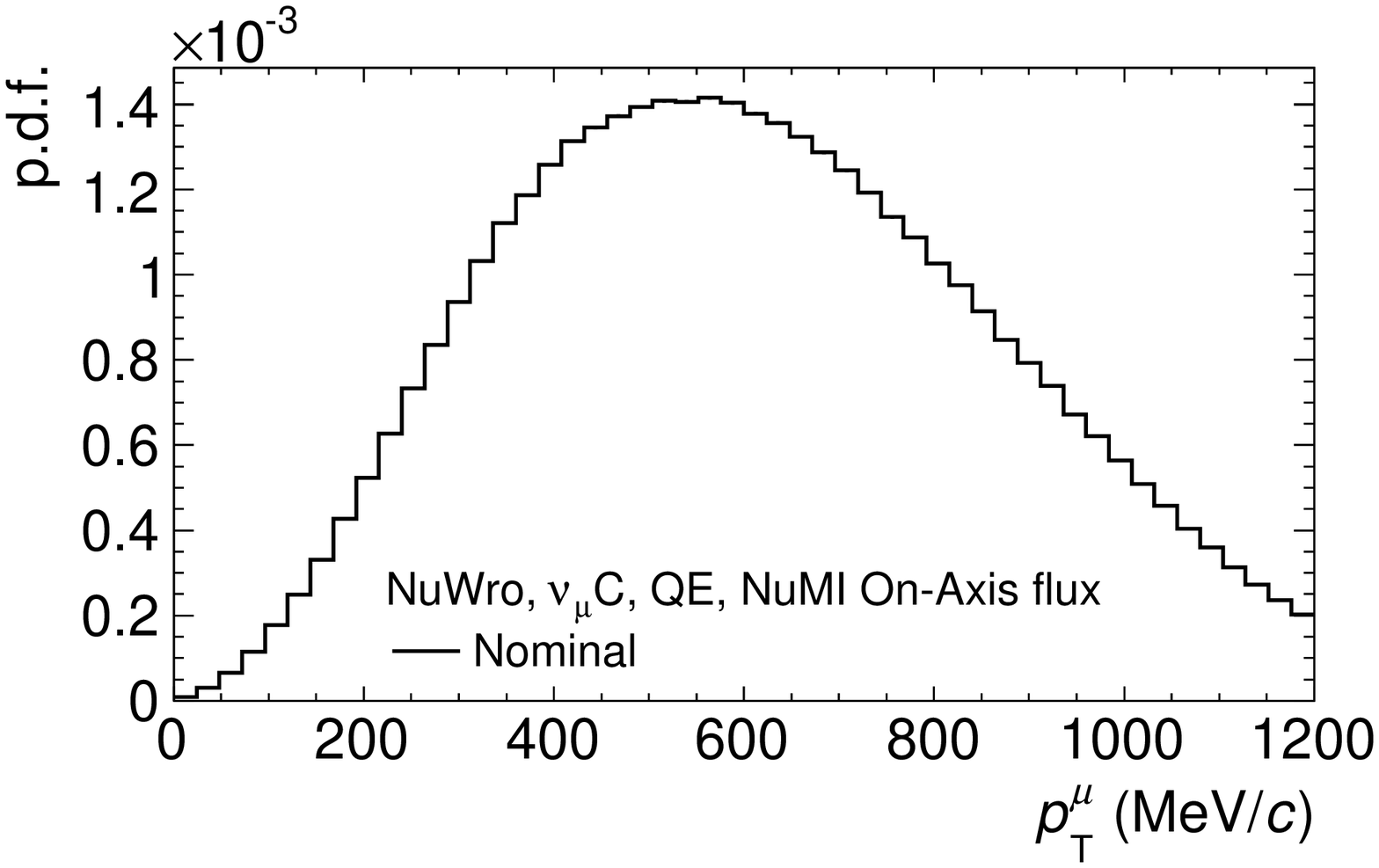}
\includegraphics[width=0.49\textwidth]{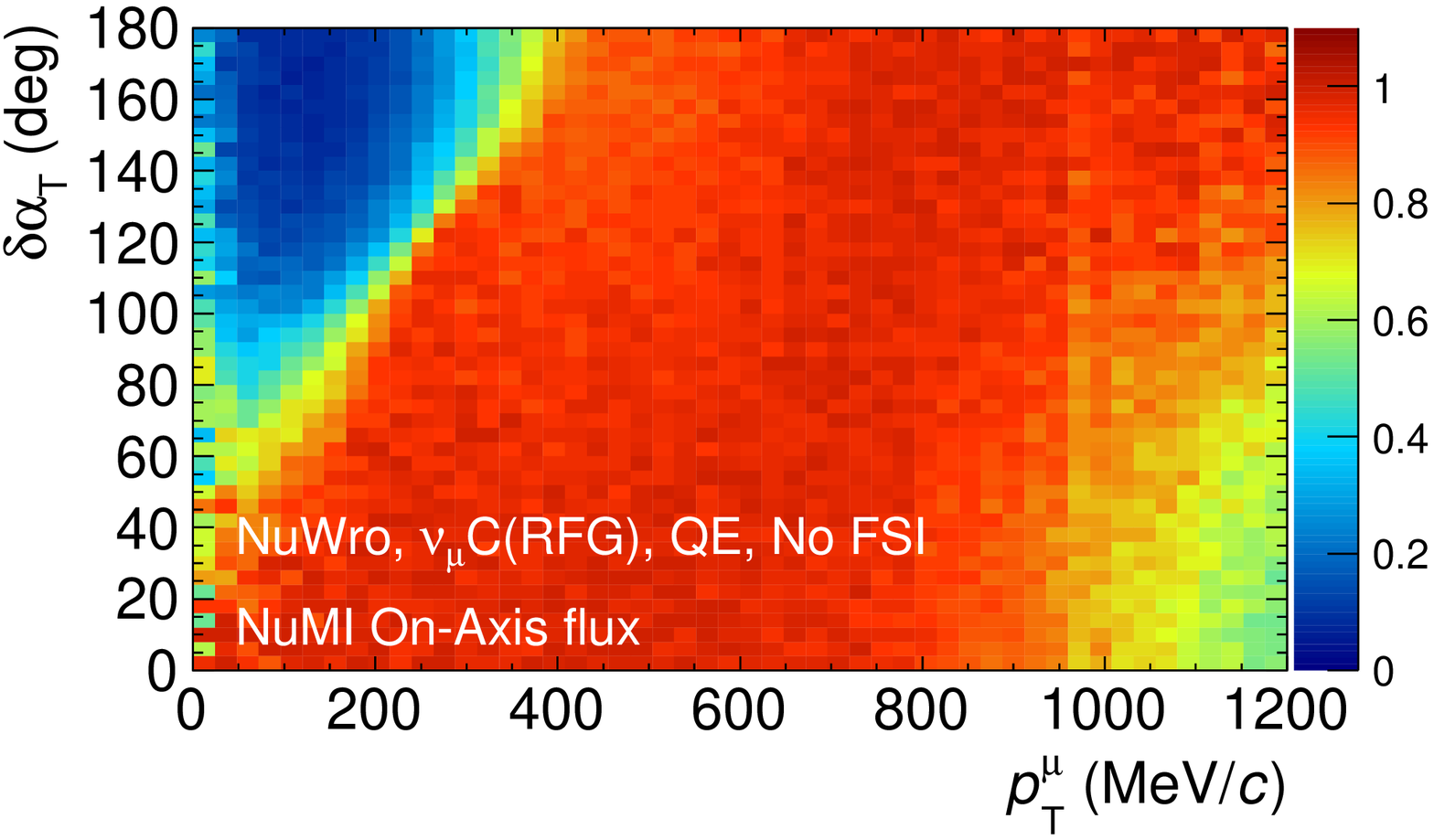}
\includegraphics[width=0.49\textwidth]{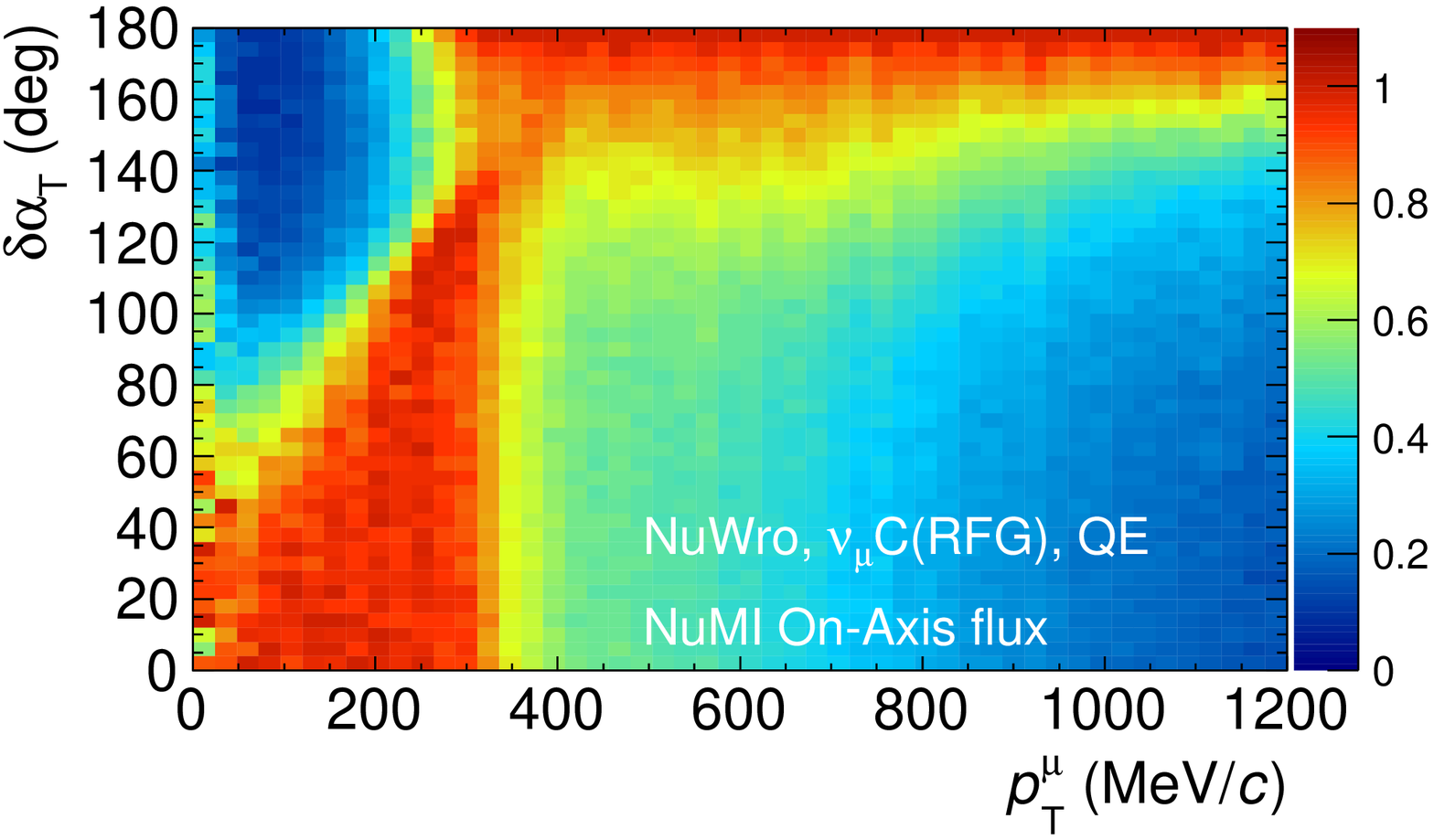}
\caption{(\textit{top}) The NuWro predicted \tmupt event rate for QE events on
carbon in the NuMi On-Axis beam.
\tmupt slice-normalised \tdat v.s \tmupt distributions, without (\textit{bottom left}), and with (\textit{bottom right}), FSIs enabled.
FSIs cause the final state hadronic system to lose momentum as it leaves the
nucleus, resulting in more events with large \tdat.}
\label{fig:datmupt}
\end{figure}

\figtxt\ref{fig:datmupt} shows how \tdat varies with \tmupt with and without FSIs
enabled in the simulation.
The transverse momentum of the charged lepton is correlated with the pre-FSI
transverse momentum of the final state hadron, up to $\textrm{k}_\textrm{f}$.
Measuring transverse kinematic imbalance as a function of the lepton momentum
may allow new insight into FSI processes.
Such a measurement would be complementary to the thin target hadron scattering
data typically used to tune FSI simulations, e.g. \cite[\S 2.5]{genieuserman}.
It may be the case that intra- and extra-nuclear forces are considerably
different, making an in-channel FSI constraint invaluable.

\subsubsection{Pauli Blocking}

\begin{figure}
\centering
\begin{minipage}[]{0.3\textwidth}
\includegraphics[width=\textwidth]{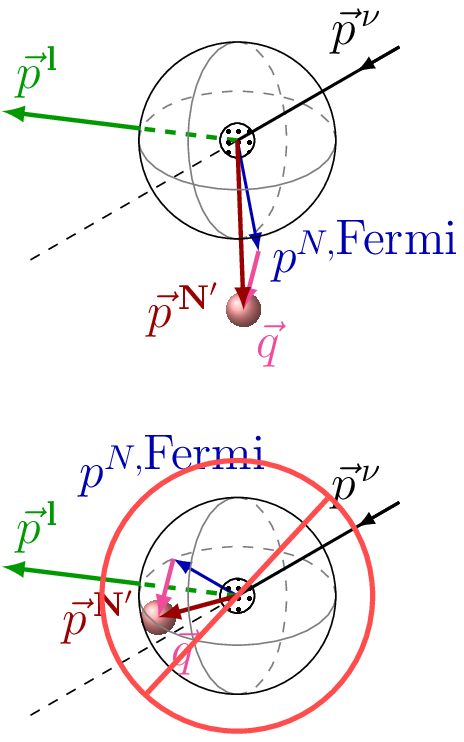}
\end{minipage}
\begin{minipage}[]{0.55\textwidth}
\includegraphics[width=0.92\textwidth]{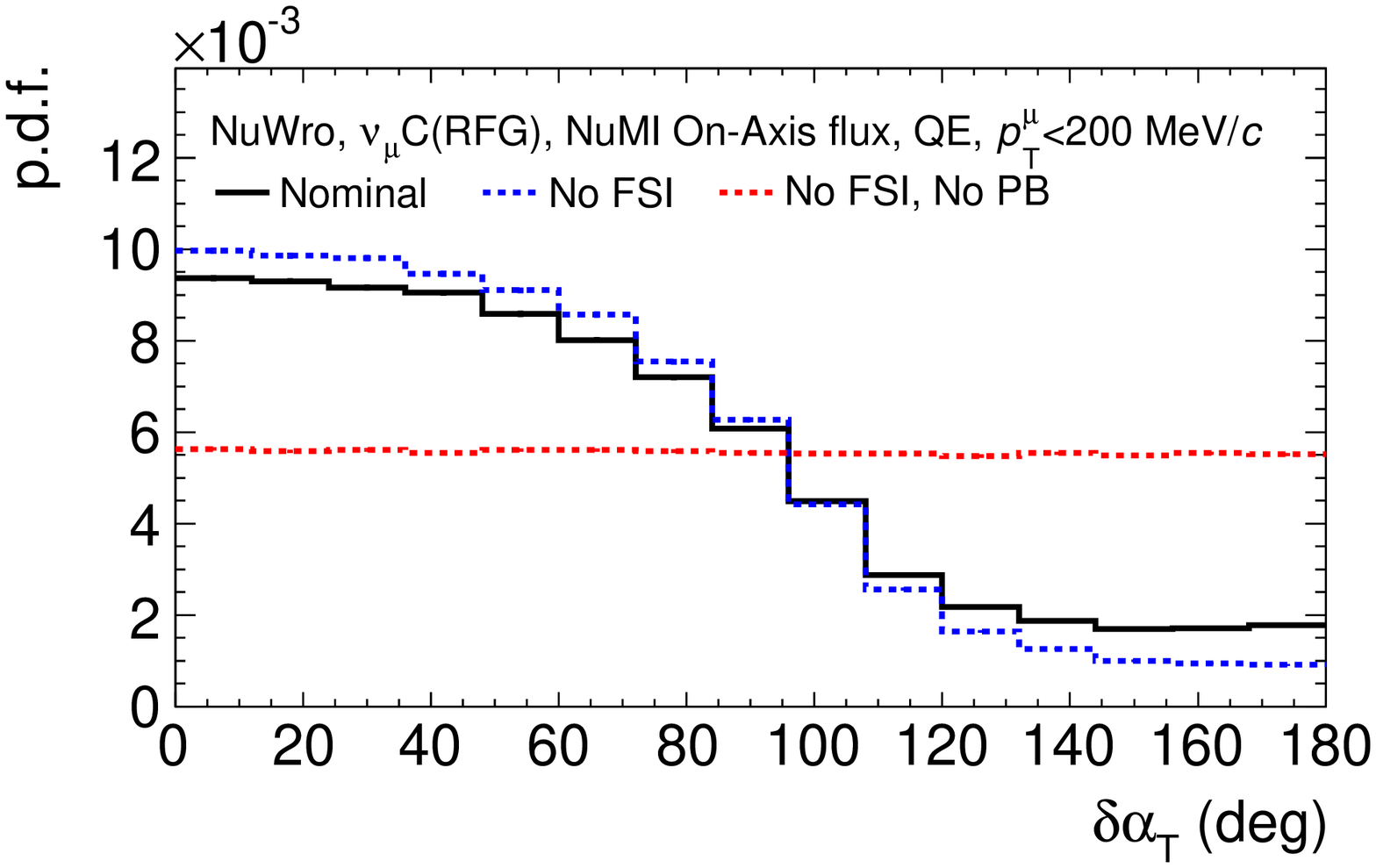}
\includegraphics[width=\textwidth]{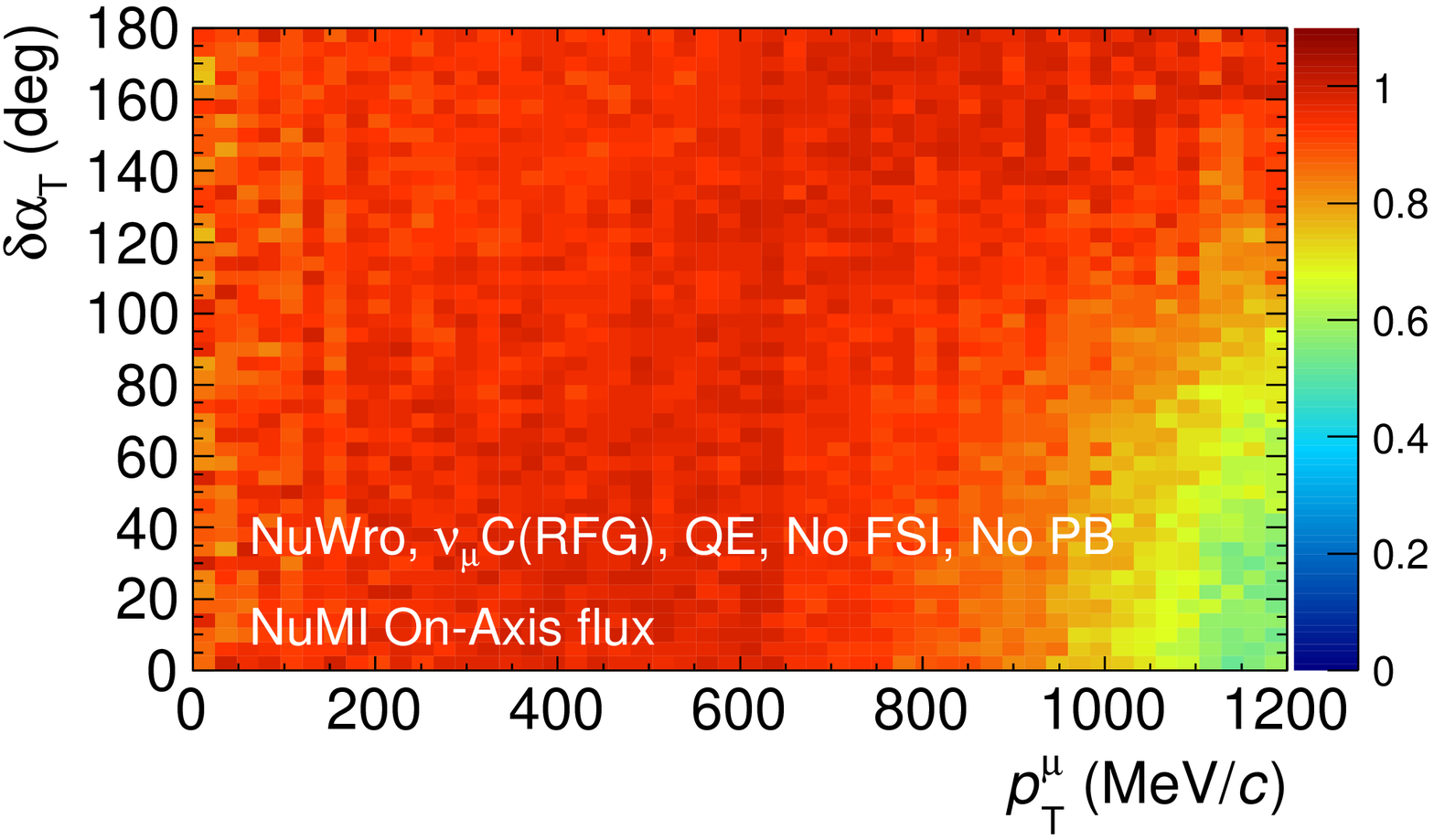}
\end{minipage}

\caption{(\textit{left}): Pauli blocking (PB) suppresses events at low \tmupt and
high \tdat because in this region a low three momentum transfer would act
against the nucleon Fermi motion leaving it in a disallowed final state.
(\textit{top right}): The high \tdat event suppression is due to Pauli blocking
and acts against the effects of FSIs at low \tmupt.
(\textit{below right}): \tmupt slice-normalised event rate in \tdat and \tmupt.
With Pauli blocking effects and FSIs removed from the simulation \tdat is very
flat in \tmupt.}
\label{fig:pauliblocking}
\end{figure}

Pauli blocking reduces the interaction cross section by limiting the phase
space for final state nucleon production.
Values of \tqsq which would result in a final state nucleon in a quantum
mechanical state that is already `filled', are disallowed.
For the global RFG nuclear model this corresponds to
$p^p \leq \textrm{k}_\textrm{f}$.
Comparison between \figtxt\ref{fig:datmupt} (\textit{left}) and
\figtxt\ref{fig:pauliblocking} (\textit{below left}) shows that Pauli blocking
causes a suppression at $\dat \sim 180^\circ$ for
$p^\mu_T \lesssim \textrm{k}_\textrm{f}$.
This region corresponds to events where a small three momentum transfer acts
against the initial state nucleon's Fermi motion, resulting in a final state
hadron with momentum below the Fermi surface---causing Pauli blocking.
Pauli blocking is also implemented in other nuclear models and the predicted
effect is qualitatively similar.

\section{Transverse Variables in Resonant Charged Pion Production}

\begin{figure}
\centering
\includegraphics[width=0.45\textwidth]{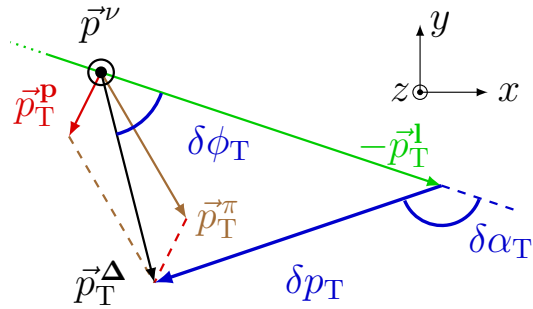}
\caption{A schematic definition of three transverse variables in a
resonant single pion production interaction.}
\label{fig:RES_stvdef}
\end{figure}

So far all discussions have been framed in the context of QE events.
However, the resonance production system is expected to be similarly balanced
for free nucleon targets.
\figtxt\ref{fig:RES_stvdef} shows the definitions of the transverse
imbalances for a $1\mu + 1p + 1\pi$ final states.
The impact of nuclear effects on transverse imbalance in RES interactions may
differ from QE.
FSIs of both protons and charged pions will be evident in these distributions.
Pauli blocking will reduce the phase space for the resonance decay, but not
limit the four momentum transfer phase space of the delta resonance production.

\begin{figure}
\centering
\includegraphics[width=0.45\textwidth]{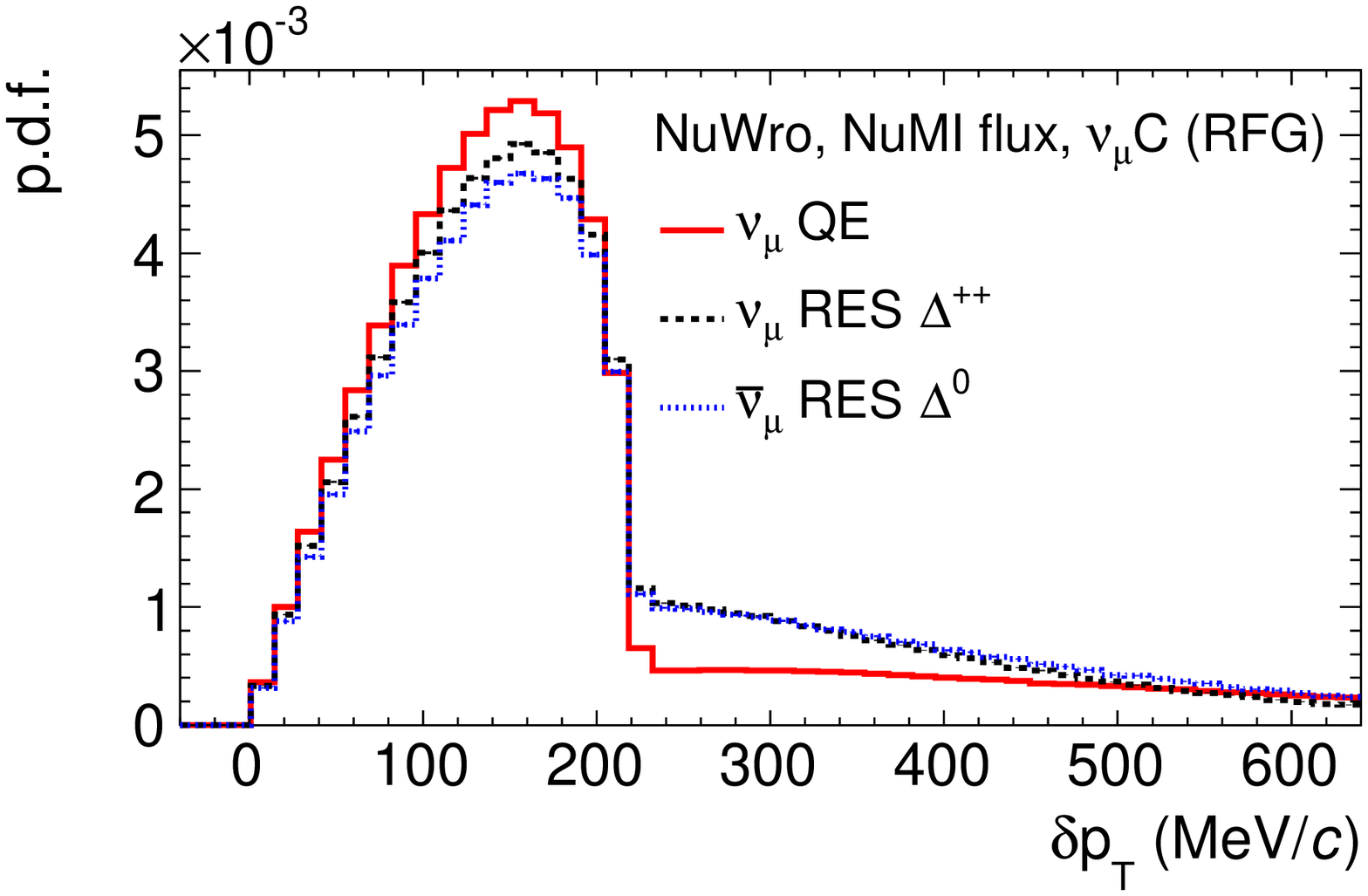}
\includegraphics[width=0.45\textwidth]{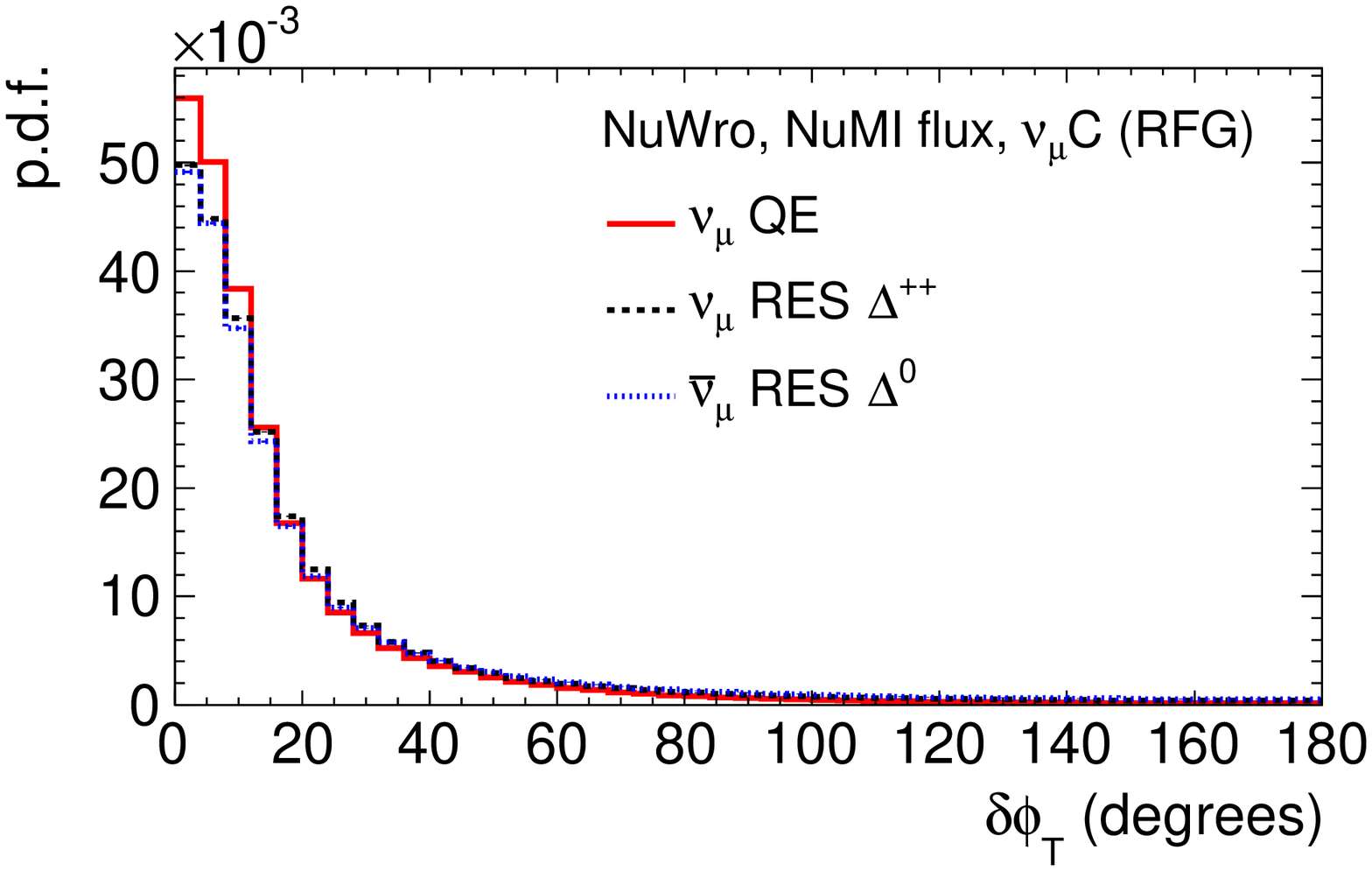}
\includegraphics[width=0.45\textwidth]{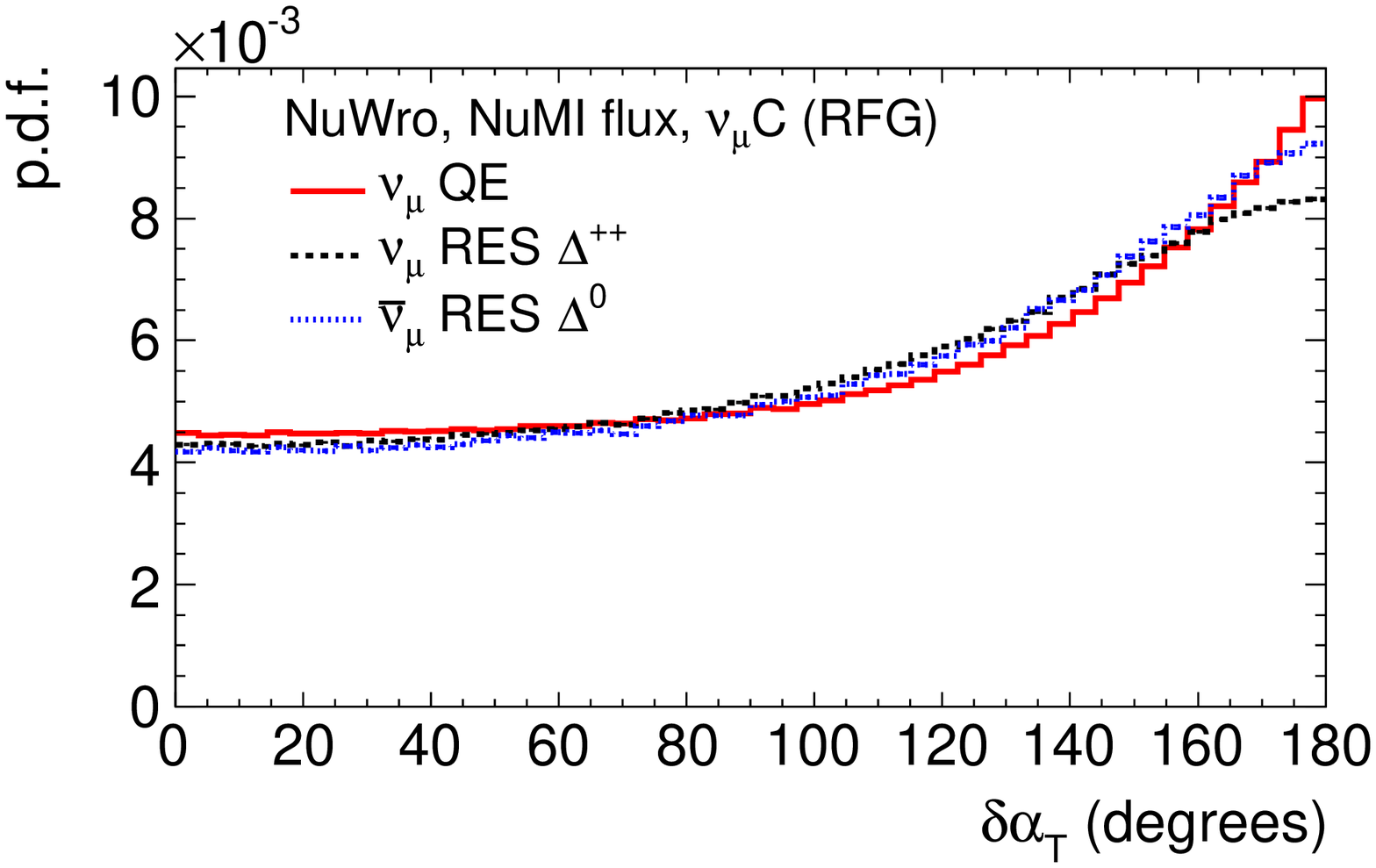}
\caption{The predicted transverse imbalance for QE, $\Delta^{++}$, and
$\Delta^{0}$ resonances.
The effect of charge pion FSI can be seen most apparently as an increased
number of events in the FSI-dominated region of \tdpt.
Measurements of these distributions will act as direct probes of proton and
proton + charged pion FSIs.}
\label{fig:RESdistribs}
\end{figure}

For the QE case it is not simple to examine transverse imbalance in an
anti-neutrino beam: the final state hadron is neutral, so accurate detection
becomes problematic.
However, possible resonance production interactions, such as
$\bar{\nu}_\ell + p \rightarrow \ell^+ + \Delta^0 \rightarrow \ell^+ p + \pi^-$,
produces only charged final state hadrons.
It is therefore possible to compare transverse imbalance across three
channels: QE, $\Delta^{++}$, and $\Delta^0$, with three distinct final states.
\figtxt\ref{fig:RESdistribs} shows the fully simulated transverse imbalance distributions for these three interaction channels.

\section{Challenges}
The transverse observable approach outlined aboves requires specific final
states, whichy may differ from the original neutrino interaction.
However, it provides a number of unique benefits.
As discussed, it is possible to make these measurements with a minimal degree of
neutrino energy dependence.
The distributions are not model dependent in their reconstruction, unlike
\tqsq and \tetrans that require assumptions on the neutrino-nucleus interaction
vertex.
Measuring transverse imbalance in neutrino scattering highlights the effects of
FSIs on neutrino interactions, rather than inferring from the results of
thin-target hadron scattering data.
The distributions are constructed from leptonic and hadronic information, so
will be naturally exclusive measurements.
This means they will be a powerful test of the full predictions of interaction
and FSI models.

Comparisons between nuclear effect-induced transverse imbalance in the three
neutrino interaction channels will give insight into intra-nuclear proton and
charged pion interactions.
This is one of a number of complementary approaches we think is important for
tackling this highly convoluted problem.

\section{Summary}

The hadronic portion of the final state is vital for identifying and
understanding neutrino interactions.
As discussed, nuclear effects obscure the primary neutrino interaction by
modifying the observed final states.
A number of the effects of using nuclear targets have been discussed in this
talk and the impact on a set of transverse variables have been shown.
Measurements of the distributions presented will provide a powerful way to gain
insight into nuclear effects, with a reduced dependence on neutrino energy.

\section*{Acknowledgments}

The author would like to thank X-G. Lu, and S. Dolan for their continued
collaboration throughout this work, K. McFarland, C. Wilkinson, C. V. C. Wret,
and Y. Hayato for various insightful discussions, and the NuWro, GENIE, and
GiBUU collaborations for their wonderful code and quickly offered assistance in
correctly using it.

\appendix
\section{NuMI Flux}\label{sec:numiflux}
\begin{figure}
\centering
\includegraphics[width=0.45\textwidth]{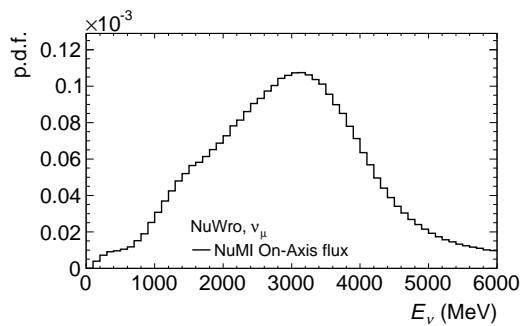}
\caption{The NuMI flux shape, as obtained from the NuWro source.}
\end{figure}
\section{Code}
During this work it was advantageous to examine the MC generator output in an
as-generator-agnostic way as possible.
To this end, code was written to allow conversion from the native GiBUU and
NEUT outputs to the so-called \texttt{rootracker} output---both NuWro and GENIE
already come with bundled software that can output in this format.
Further more the \texttt{rootracker} event vectors were further processed into
\texttt{ROOT}-based analysis files.
All of the code is open source and available upon request.
Discussions of designing a better event format than \texttt{rootracker} are
very welcome.

\end{document}